\documentclass[aps,twocolumn,floatfix,showpacs,amsmath,amsfonts]{revtex4}

\usepackage{graphicx}
\usepackage{bm}

\newcommand{\y}{{\bf y}}
\newcommand{\x}{{\bf x}}
\newcommand{\p}{{\bf p}}
\newcommand{\q}{{\bf q}}

\newcommand{\g}{\gamma}
\newcommand{\uu}{{\bf u}}

\begin{document}

\title{Fictitious time wave packet dynamics:
I. Nondispersive wave packets in the quantum Coulomb problem}

\author{Toma\v{z} Fab\v{c}i\v{c}}
%\email{fabcic@itp1.uni-stuttgart.de}
\author{J\"org Main}
%\email{main@itp1.uni-stuttgart.de}
\author{G\"unter Wunner}
%\email{wunner@itp1.uni-stuttgart.de}
\affiliation{Institut f\"ur Theoretische Physik 1, Universit\"at Stuttgart,
  70550 Stuttgart, Germany}
\date{\today}

\begin{abstract}
Nondispersive wave packets in a fictitious time variable are calculated
analytically for the field-free hydrogen atom.
As is well known by means of the Kustaanheimo-Stiefel transformation the 
Coulomb problem can be converted into that of a four-dimensional harmonic 
oscillator, subject to a constraint. This regularization makes use of a 
fictitious time variable, but  arbitrary Gaussian wave packets in that 
time variable in general violate that constraint.
The set of ``restricted Gaussian wave packets'' consistent with the constraint
is constructed and shown to provide a complete basis for the expansion of
states in the original three-dimensional coordinate space.
Using that expansion arbitrary localized Gaussian wave packets of the
hydrogen atom can be propagated analytically, and exhibit a nondispersive
periodic behavior as functions of the fictitious time.
Restricted wave packets with and without well defined angular momentum quantum numbers
are constructed.
They will be used as trial functions in time-dependent variational
computations for the hydrogen atom in static external fields in the 
subsequent paper [T.\ Fab\v{c}i\v{c} et al., submitted].
\end{abstract}

%\keywords{}

\maketitle

\section{Introduction}
\label{sec:intro}
Wave packets play an important role in the description of atoms, e.g., 
for the understanding of ionization processes in microwave experiments 
\cite{Bay74,Gal88} or in experiments with short laser pulses \cite{Jon93,Jon96}.
Theoretically, the wave packet propagation can be calculated by exact
quantum computations \cite{Par86} or approximately with, e.g., 
semiclassical \cite{Alb99} or variational \cite{Hor92} techniques.
Contrary to the harmonic oscillator, where Gaussian wave packets and
coherent states \cite{Schroedinger26} can easily be described analytically,
the evolution of arbitrary Rydberg wave packets is nontrivial already in the 
pure Coulomb problem where spreading and revival phenomena are observed in 
the long-time computer simulation of a quantum wave packet \cite{Str93}.
Coherent states for the hydrogen atom have been constructed by Klauder
\cite{Kla96} and by Majumdar and Sharatchandra \cite{Maj97}, however, 
Bellomo and Stroud \cite{Bel98,Bel99} have shown that these states do
not move quasiclassically but spread rapidly over the Keplerian orbit.
Dispersion is a general property of Rydberg wave packets with the exception 
of nondispersive electronic wave packets existing in periodically driven 
atoms such as the hydrogen atom in microwave fields \cite{Buch95,Cer97}.

The success of applying variational methods to wave packet propagation
crucially depends on the choice of the trial function.
Gaussian wave packets (GWPs) are certainly well suited for smooth and 
nearly harmonic potentials \cite{Hel75,Hel76}.
The Coulomb potential is not a promising candidate for successfully 
propagating GWPs directly.
Nevertheless, the GWP method based on the local harmonic approximation has 
been applied in one dimension to the singular Coulomb potential 
\cite{Barnes93,Barnes94,Barnes95}. 
In the three-dimensional space a regularization in Kustaanheimo-Stiefel (KS) 
coordinates \cite{KuS65,Stiefel71}
originally introduced for the Kepler problem in classical celestial mechanics, 
but also adapted to the hydrogen atom \cite{Boit72}, transforms the Coulomb 
potential to a harmonic potential with a constraint.
In the regularized hydrogen atom the application of the GWP method should 
therefore be capable of yielding exact results when the constraint can be 
handled. 
The regularization implies a fictitious time variable which has been shown 
to be the eccentric anomaly of the corresponding classical orbit 
\cite{Johnson87}.
Various approaches have been made to construct coherent states for the 
hydrogen atom in the fictitious time 
\cite{Ger86,Ger88,Tad98,Xu00,Una01,Polshin01,Gur04} 
in analogy with the coherent states of the harmonic oscillator.
These approaches construct the coherent states as the eigenstates of the 
lowering operators associated with the harmonic potential.

In this paper we consider the field-free hydrogen atom and show that 
contrary to the dynamics in the real physical time the exact propagation 
of arbitrary initial Gaussian wave packets in the fictitious time can be 
described analytically and exhibit a nondispersive periodic time dependence.
In Sec.\ \ref{sec_reg} the Coulomb problem is transformed to the problem of the
four-dimensional (4D) harmonic oscillator in Kustaanheimo-Stiefel coordinates
 subject to  a constraint.
The consequences of the constraint for Gaussian wave packets are discussed
in Sec.\ \ref{sec_4D_GWP} and the physically allowed set of ``restricted 
Gaussian wave packets'' is constructed.
In Sec.\ \ref{sec_analytic_prop} the analytical time evolution is derived 
for initially three-dimensional (3D) Gaussian wave packets in the physical 
space, and also for two-dimensional (2D) and one-dimensional (1D) wave 
packets with cylindrical and spherical symmetry, respectively.

In the subsequent paper \cite{Fab08b} the investigations are extended from
the pure Coulomb problem to the hydrogen atom in static external electric
and magnetic fields.
Wave packets are propagated by application of the time-dependent variational
principle in such a way that the dynamics is exact for the Coulomb problem
and approximations in the variational approach are only induced by the 
external fields.
Quantum spectra of the nonintegrable systems are then obtained by the frequency
analysis of the time autocorrelation function of the propagated wave function.

\section{Regularization of the hydrogen atom}
\label{sec_reg}
To make our presentation self-contained we briefly review the 
Kustaanheimo-Stiefel transformation for the Coulomb problem.
The time-independent Schr\"odinger equation for the hydrogen atom reads
\begin{equation}
H_3 \psi=\left( -\frac{1}{2}\Delta_3 - \frac{1}{r}\right) \psi = E \psi, 
\label{eq_h_3d}
\end{equation}
with $\Delta_3$ the Cartesian form of the Laplacian.
A regularization of the singular Coulomb potential is obtained by using
Kustaanheimo-Stiefel (KS) coordinates $\uu = (u_1,u_2,u_3,u_4)$ 
\cite{KuS65,Stiefel71} which are introduced here, according to Ref.\ 
\cite{Ger86}, 
differing by a factor of two from the original definition,
\begin{eqnarray}
 x & = & u_1 u_3 - u_2 u_4, \nonumber \\
 y & = & u_1 u_4 + u_2 u_3, \nonumber \\
 z & = & \frac{1}{2}\left( u_1^2+u_2^2-u_3^2-u_4^2 \right).
\label{KScoo}
\end{eqnarray} 
By adding a fourth component with the constant value zero to the 
physical position vector, i.e.\ $\x=(x,y,z,0)$, the transformation can be 
written in matrix notation,
\begin{equation}
 \x =L({\bf u}) {\bf u} \; ,
\end{equation}
with 
\begin{equation}
L({\bf u})=\frac{1}{2}\left(\begin{array}{crrr}
 u_3 & -u_4 &  u_1 & -u_2  \\
 u_4 &  u_3 &  u_2 &  u_1  \\
 u_1 &  u_2 & -u_3 & -u_4  \\
 u_2 & -u_1 & -u_4 &  u_3  
 \end{array} \right) \; .
\end{equation}
The introduction of the auxiliary degree of freedom, which renders the 
originally three-dimensional problem four-dimensional, entails a constraint
on physically allowed wave functions $\psi$, i.e.\
\begin{equation}
 X \psi \equiv
 \left(u_2\frac{\partial}{\partial u_1} - u_1\frac{\partial}{\partial u_2}
 - u_4\frac{\partial}{\partial u_3}+ u_3\frac{\partial}{\partial u_4}\right) 
 \psi = 0 \; .
\label{KScon}
\end{equation}
With $r=(x^2+y^2+z^2)^{1/2}={\bf u}^2/2$ the Schr\"odinger equation 
\eqref{eq_h_3d} transformed in Kustaanheimo-Stiefel coordinates reads
\begin{equation}
 \left(-\frac{1}{2 \uu^2}\Delta_4 - \frac{2}{\uu^2}\right) \psi = E \psi \; ,
\end{equation}
where $\Delta_4$ denotes the 4D Cartesian form of the Laplacian.
Multiplication with $\uu^2$ and reordering of the terms yields
\begin{equation}
 H \psi=\left(-\frac{1}{2}\Delta_4 - E\uu^2\right) \psi  = 2 \psi \; .
\label{eq_zw1}
\end{equation}
Eq.\ \eqref{eq_zw1} is not a standard linear eigenvalue problem.
Scaling the coordinates 
\begin{equation}
 \uu \rightarrow {\sqrt{n}}\, \uu \; , \qquad  H \rightarrow n \, H \; ,
\label{scal}
\end{equation}
and setting 
\begin{equation}
 E = -\frac{1}{2n^2} \; ,
\label{energy_eigenvalues}
\end{equation}
leads to the time-independent Schr\"odinger equation 
\begin{equation}
 H\psi =\left(-\frac{1}{2} \Delta_4 + \frac{1}{2}\uu^2\right)\psi = 2n\psi \; ,
\label{H_KSreg}
\end{equation}
which represents the Schr\"odinger equation of the 4D harmonic 
oscillator subject to the constraint \eqref{KScon}.
The scaling parameter,  which takes only integer values $n=1,2,3,\dots$, 
turns out to be the principal quantum number of the hydrogen atom. 
The 4D isotropic harmonic oscillator is invariant under the
unitary group $U(4)$ and thus the eigenstates of Eq.\ \eqref{H_KSreg} are 
not unique.
The simple product of four eigenstates of the 1D harmonic 
oscillators, i.e., the separation of Eq.\ \eqref{H_KSreg} in the Cartesian 
coordinates $(u_1,u_2,u_3,u_4)$, in general violates the constraint 
\eqref{KScon} and therefore represents unphysical solutions.
The constraint \eqref{KScon} rather suggests the introduction of 
two sets of polar coordinates, viz.\ the semiparabolic coordinates
\begin{eqnarray}
 u_1 &=& \mu \cos \varphi_\mu \; , \quad
 u_2 = \mu \sin \varphi_\mu \; , \nonumber \\
 u_3 &=& \nu \cos \varphi_\nu \; , \quad
 u_4 = \nu \sin \varphi_\nu \; , 
\label{semipara}
\end{eqnarray}
with the associated angular momenta 
($p_j=\frac{1}{i} \frac{\partial}{\partial u_j}$)
\begin{equation}
\begin{array}{ccccc}
L_\mu & = u_1p_2-u_2p_1 & = & \frac{1}{i}\frac{\partial}{\partial\varphi_\mu}, \\
L_\nu & = u_3p_4-u_4p_3 & = & \frac{1}{i}\frac{\partial}{\partial\varphi_\nu}.
\end{array}
\label{Lphi}
\end{equation}
Eq.\ \eqref{Lphi} yields the constraint \eqref{KScon} in the form
\begin{equation}
L_\mu=L_\nu \equiv L_z.
\end{equation}
The relation between the physical Cartesian coordinates and the 
semiparabolic coordinates is obtained using the definitions \eqref{KScoo} 
and \eqref{semipara}
\begin{eqnarray}
 x &=& \mu\nu (\cos\varphi_\mu\cos\varphi_\nu - \sin\varphi_\mu\sin\varphi_\nu)
 \nonumber \\
   &=& \mu\nu\cos (\varphi_\mu + \varphi_\nu) = \mu\nu\cos\varphi \; ,
 \nonumber \\
 y &=& \mu\nu (\cos\varphi_\mu\sin\varphi_\nu + \sin\varphi_\mu\cos\varphi_\nu)
 \nonumber \\
   &=& \mu\nu\sin (\varphi_\mu + \varphi_\nu) = \mu\nu\sin\varphi \; ,
 \nonumber \\
 z &=& \frac{1}{2} (\mu^2 - \nu^2) \; ,
\end{eqnarray}
with the physical azimuthal angle $\varphi = \varphi_\mu + \varphi_\nu$.
In the semiparabolic coordinates the Schr\"odinger equation reads
\begin{equation}
 \left[ -\frac{1}{2} \Delta_\mu- \frac{1}{2} \Delta_\nu + \frac{1}{2}
 \left( \mu^2 + \nu^2 \right) \right] \psi = 2n\; \psi \; , 
\label{eq_semi_para_reg_h}
\end{equation} 
with
\begin{equation}
 \Delta_\rho = \frac{1}{\rho}\frac{\partial}{\partial \rho} \rho
 \frac{\partial }{\partial \rho}+ \frac{1}{\rho^2} \frac{\partial^2 }
 {\partial \varphi^2} \; , \quad \rho = \mu, \nu \; .
\label{eq_Laplace_op_semi}
\end{equation}
Eq.\ \eqref{eq_semi_para_reg_h} is separated in two uncoupled 2D harmonic 
oscillators 
in the coordinates $\mu,\varphi_\mu$ and $\nu, \varphi_\nu$, respectively.
The solution can be taken in the product form
\begin{equation}
\psi(\mu,\nu,\varphi) =  \Phi_{N_\mu m}(\mu)\Phi_{N_\nu m}(\nu)e^{im\varphi}
\end{equation}
where 
\begin{gather}
 \left[-\frac{1}{2 \rho}\frac{\partial}{\partial \rho} \rho
 \frac{\partial }{\partial \rho}+ \frac{m^2}{2 \rho^2} + \frac{1}{2}\rho^2
 \right] \Phi_{N_\rho m}(\rho) \nonumber \\
 = (2N_\rho+ |m|+1)\Phi_{N_\rho m}(\rho) \; ,
\label{eq_laplc_munu}
\end{gather}
with $\rho = \mu,\nu$ and $N_\rho = 0,1,2,\dots$.
The coordinate representation of the eigenstates is
\begin{equation}
 \Phi_{N_\rho m}(\rho) = \sqrt{\frac{N!}{\pi(N_\rho+|m|)}}
 \rho^{|m|}L_{N_\rho}^{|m|}(\rho^2) e^{-\frac{1}{2} \rho^2} \; ,
\label{eq_eigen_f}
\end{equation}
with the associated Laguerre polynomials $L_{N_\rho}^{|m|}$.
For the principal quantum number $n$ introduced above we obtain the relation 
\begin{equation}
 n = N_\mu + N_\nu + |m|+1 = 1,2,3,\dots \; ,
\label{eq_eigenv_2D}
\end{equation}
and therefore via Eq.\ \eqref{energy_eigenvalues} the correct Rydberg spectrum.

To perform time-dependent computations it is necessary to formulate 
the time-dependent version of the Schr\"odinger equation \eqref{H_KSreg}. 
By analogy with the usual identification 
$E \rightarrow i\frac{\partial}{\partial t}$ where $t$ is the physical time, 
the ``fictitious time'' variable $\tau$ is introduced, 
 (see, e.g., Refs.\ \cite{Ger86,Tad98})
as the conjugate variable to the principal quantum number $n$,
\begin{equation}
 2n \rightarrow i \frac{\partial}{\partial \tau} \; .
\label{eq_fict_time}
\end{equation} 
The regularized Schr\"odinger equation for the hydrogen atom in the 
fictitious time then reads
\begin{equation}
 i \frac{\partial}{\partial\tau} \psi = H \psi 
 = \left(-\frac{1}{2} \Delta_4 + \frac{1}{2} \uu^2\right) \psi \; .
\label{eq_regH}
\end{equation}
In the following the ``fictitious time'' $\tau$ will simply be denoted as
``time'' for brevity, whereas $t$ will be named ``physical time''.

The form of Eq.\ \eqref{eq_regH} suggests that it can simply be solved by 
Gaussian wave packets
in Kustaanheimo-Stiefel coordinates $\uu$, i.e.\ 
\begin{equation}
 g({\bf y},\uu)=e^{i[(\uu-\q)A(\uu-\q)+{\boldsymbol \pi}\cdot(\uu-\q)+\gamma]} \; ,
\label{GWPhel1}
\end{equation}
where $A$ is a complex symmetric $4\times 4$ matrix with positive definite 
imaginary part and the momentum ${\boldsymbol \pi}$ and the 
center $\q$ are real, 
 4D vectors in the Kustaanheimo-Stiefel coordinates. Those vectors  
 represent  
the expectation values of the position and the momentum operator, 
respectively, i.e.\
$\q=\langle g|\uu|g \rangle / \langle g|g \rangle$ and 
${\boldsymbol\pi} =\langle g|\frac{1}{i}\nabla_4|g\rangle/ \langle g | g \rangle$. 
The phase and normalization is given by the complex scalar $\g$. 
Collectively the parameters are denoted by 
${\bf y}=(A,{\boldsymbol\pi},\q,\g)$, which is a set of
$4\times(4+1)/2+4+1=15$ complex parameters when the two real vectors 
$\q$ and ${\boldsymbol\pi}$ are counted as a single complex vector.
Inserting \eqref{GWPhel1} into the Schr\"odinger equation \eqref{eq_regH}
yields a set of ordinary differential equations for the parameters ${\bf y}$,
which can be solved analytically for any given initial GWP.
As the potential is harmonic the time-dependent GWPs are exact solutions of
Eq.\ \eqref{eq_regH}.

The problem, however, is that in Kustaanheimo-Stiefel coordinates the 
constraint \eqref{KScon} on 
physically allowed wave functions must be taken into account.
 The GWPs \eqref{GWPhel1} in general violate that constraint.
The question is whether the constraint can be fulfilled exactly by a 
Gaussian at all, and, if so, whether GWPs fulfilling 
the constraint are reasonable trial functions in the sense that they 
still present a complete basis set.

The advantage of the formulation of the Hamiltonian in semiparabolic 
coordinates \eqref{eq_semi_para_reg_h} is that the constraint \eqref{KScon}
is already incorporated.
However, in semiparabolic coordinates it is nontrivial to find Gaussian type 
trial functions for the exact solution of the time-dependent Schr\"odinger 
equation because the Laplacian is not of Cartesian form, see Eq.\  
\eqref{eq_Laplace_op_semi}, and includes centrifugal barriers.
In the following we will therefore use the Cartesian type KS coordinates
rather than semiparabolic coordinates to investigate the impact of the 
constraint \eqref{KScon} on a Gaussian trial function and to discuss the 
properties of the resulting ``restricted Gaussian wave packets''.

\section{Restricted Gaussian wave packets}
\label{sec_4D_GWP}
The regularization of the hydrogen atom in Sec.\ \ref{sec_reg} has 
transformed the Coulomb potential to a harmonic potential in the 
Schr\"odinger equation \eqref{H_KSreg}.
The goal now is to perform exact wave packet propagation in the fictitious time 
$\tau$ for the hydrogen atom. The 
constraint can be fulfilled by a 4D GWP if the space of admissible 
configurations of the parameters $\y$ is restricted. 
The structure matrix
\begin{equation} 
 J=\left(\begin{array}{rccr}
  0 & 1 & 0 &  0 \\
 -1 & 0 & 0 &  0 \\
  0 & 0 & 0 & -1 \\
  0 & 0 & 1 &  0 
\end{array}\right) \; ,
\end{equation} 
allows for the compact notation of the constraint \eqref{KScon},
\begin{equation}
 X \psi = \uu J^T \nabla_4 \psi = 0 \; .
\end{equation}
Letting the constraint operator $X$ act on the trial function 
\eqref{GWPhel1} yields
\begin{equation}
 X g({\bf y},\uu)
 = \uu J^T [2A(\uu-\q)+ {\boldsymbol\pi}]\;g({\bf y},\uu) \overset{!}{=}0 \; .
\label{KScon_gwp}
\end{equation}
Thus the result is a quadratic polynomial in the coordinates $\uu$, 
 multiplied by the GWP itself.
The constraint \eqref{KScon_gwp} has to be satisfied pointwise for all 
$\uu \in {\mathbb R}^4$. 
For nontrivial wave packets the polynomial in \eqref{KScon_gwp} must vanish, 
and an algebraic equation remains,
\begin{equation}
 2 \uu J^T  A \uu +   \uu J^T({\boldsymbol \pi}-2A\q) \overset{!}{=} 0 \; ,
\label{eq_sec_ord_poly}
\end{equation}
which is only possible if all coefficients of the second order polynomial 
in \eqref{eq_sec_ord_poly} are zero. 
Let us first investigate the term linear in ${\bf u}$ whose coefficients 
must vanish, i.e.\ ${\boldsymbol\pi}=2A\q$.
Inserting this condition into the wave function \eqref{GWPhel1} yields
$g({\bf y},\uu)=\exp\{i[(\uu-\q)A(\uu+\q)+\gamma]\}
=\exp\{i[\uu A\uu+\gamma']\}$ with $\gamma'=\gamma-\q A\q$, which means
that without loss of the variational freedom we can set
\begin{equation}
 \q = 0 \; , \quad \mbox{and} \quad {\boldsymbol \pi} = 0 \; ,
\label{eq_nb_0}
\end{equation}
because any nonzero $\q$ and ${\boldsymbol\pi}$ vectors only change the 
scalar $\gamma$.
The bilinear form in \eqref{eq_sec_ord_poly} must also be zero, i.e.\
$2 \uu J^T A \uu = 0$.
This requires the matrix of the bilinear form 
\begin{equation*}
 J^TA = \left(
\begin{array}{rrrr}
 -a_{12} & -a_{22} & -a_{23} & -a_{24} \\
  a_{11} & a_{12} & a_{13} & a_{14} \\
  a_{14} & a_{24} & a_{34} & a_{44} \\
 -a_{13} & -a_{23} & -a_{33} & -a_{34} 
\end{array} \right)
%\label{eq_nb_skew}
\end{equation*}
to be skew-symmetric, i.e., the diagonal elements of $J^TA$ must vanish,
$a_{12}=a_{34}=0$, and from the off diagonal elements we obtain
$a_{11}=a_{22}$, $a_{33}=a_{44}$, $a_{24}=-a_{13}$, and $a_{23}=a_{14}$.
With the definitions $a_{11}=a_\mu$, $a_{33}=a_\nu$, $a_{13}=a_x$, $a_{14}=a_y$, 
the matrix $A$ must be of the form  
\begin{equation}
 A = \left(\begin{array}{crcr}
  a_{\mu} & 0\; & a_{x} & a_{y} \\
  0      & a_{\mu} & a_{y} & -a_{x} \\
  a_{x} & a_{y} & a_{\nu} & 0\; \\
  a_{y} & -a_{x} & 0     & a_{\nu} 
\end{array}\right) \; .
\label{eq_A_sym}
\end{equation}
Eqs.\ \eqref{eq_nb_0} and \eqref{eq_A_sym} imply that the set of 15
complex parameters ${\bf y}$ of the general GWP \eqref{GWPhel1} is reduced 
to only five parameters ${\bf y}=(a_\mu,a_\nu,a_x,a_y,\gamma)$ for a 
restricted GWP satisfying the constraint \eqref{KScon}!

The question arises whether the restricted GWPs form a complete basis set, 
such that any physically allowed state can be expanded in this basis.
It is not evident that a superposition of restricted GWPs whose 
centers are all located at the origin and which only differ by their complex widths, 
is flexible enough to represent arbitrary quantum states. 
The usual form of the resolution of the identity \cite{Hub88} for a 
continuous basis set of normalized, unrestricted GWPs of the form 
\eqref{GWPhel1} is
\begin{equation}
 \frac{1}{(2\pi)^4} \int d \pi^4 dq^4
 |g(\y) \rangle \langle g(\y)| = {\boldsymbol 1} \; ,
\label{id_GWP}
\end{equation} 
where the width of each GWP basis state is kept fixed.
Obviously, Eq.\ \eqref{id_GWP} cannot be applied to the restricted GWPs, 
since both parameters ${\boldsymbol\pi}$ and $\q$ are set to zero 
(see Eq.\ \eqref{eq_nb_0}).
However, it is sufficient to require the restricted GWPs to be complete in 
the 3D physical space only.
To verify the completeness in the 3D space we transform the restricted GWP 
in KS coordinates back into the original 3D Cartesian coordinates,
\begin{subequations}
\label{eq_res_gwp}
\begin{align}
 g({\bf y},\x)
 & = e^{i(\uu A \uu + \gamma)} 
\label{eq_gwpKS} \\
 & = e^{i[a_\mu(u_1^2+u_2^2)+a_\nu(u_3^2+u_4^2)]} \nonumber \\
 & \times e^{i[2a_x(u_1 u_3-u_2 u_4)+2a_y(u_1 u_4+u_2 u_3)+\gamma]} \\
 & = e^{i(a_\mu \mu^2+a_\nu \nu^2  +2a_x x+2a_y y+\gamma)}  
\label{eq_KS_gwp} \\
 & = e^{i[(a_\mu+a_\nu) r+(a_\mu-a_\nu) z  +2a_x x+2a_y y+\gamma]} \\
 & = e^{i(p_r r+ \p \cdot \x + \gamma)} \; ,                 
\label{eq_KS_gwp_x}
\end{align}
\end{subequations}
where we have exploited that in semiparabolic coordinates 
 $\mu^2 = r+z$, $\nu^2 = r-z$.
In Eq.\ \eqref{eq_KS_gwp_x} the set of parameters $(a_\mu,a_\nu,a_x,a_y)$ has 
been
replaced by an equivalent set of complex parameters, defined by 
\begin{eqnarray}
 p_r &=& a_\mu+a_\nu \; , \nonumber \\
 {\bf p} &=& (p_x,p_y,p_z) = (2a_x,2a_y,a_\mu-a_\nu) \; .
\label{eq_aparam_cart}
\end{eqnarray}
For $p_r=0$ and real-valued parameters $p_x,p_y,p_z$ the restricted GWP in 
Cartesian coordinates \eqref{eq_KS_gwp_x} reduces to a plane wave 
$e^{i\p\cdot\x}$.
(Note that in that case the imaginary part of the matrix $A$ is not 
positive definite and the wave function cannot be normalized.)
Since plane waves are known to form a complete basis it is proved
that the restricted GWPs, forming a superset of plane waves, are also 
complete, or even over-complete.
However, they do not form a complete basis set of the 4D harmonic
oscillator \eqref{H_KSreg}.

The GWPs \eqref{eq_res_gwp} satisfy the constraint \eqref{KScon}, but
``for the price'' of the condition \eqref{eq_nb_0}, i.e., they are 
localized around the origin with zero mean velocity.
It seems impossible that the time propagation of a single restricted GWP
exhibits any meaningful dynamics with, in particular, a classical limit 
in the sense of the correspondence principle.
However, a wave packet localized in the physical coordinate and momentum
space of the hydrogen atom can be constructed as a superposition of the
restricted GWPs.
The expansion and exact time evolution of wave functions in the basis 
\eqref{eq_res_gwp}, and in addition, in modified bases sets for certain 
symmetry subspaces of the hydrogen atom are the subjects of the next section.

\section{Analytical wave packet dynamics in the hydrogen atom}
\label{sec_analytic_prop}
An arbitrary wave packet of the hydrogen atom can be propagated analytically 
in the fictitious time.
This is achieved by expanding the initial wave packet in terms of 
the restricted GWPs, whose time-dependence is derived and shown to 
be given by simple analytical formulae.
We consider three different cases.
In Sec.\ \ref{sec_expansion_res} the time propagation of Gaussian wave packets
in the 3D physical space is discussed.
These wave packets are not eigenstates of the angular momentum operator.
We then introduce the time propagation of wave packets with symmetries, viz.\
in Sec.\ \ref{sec_g_lz_m} cylindrically symmetric 2D wave packets with well 
defined magnetic quantum number $m$, and in Sec.\ \ref{sec_radial_gwp}
spherically symmetric 1D wave packets with well defined angular momentum 
quantum numbers $l$ and $m$.
\subsection{Propagation of 3D Gaussian wave packets}
\label{sec_expansion_res}
The aim is an exact time propagation of arbitrary wave functions in the 
hydrogen atom.
The initial wave function $\psi(0)$ is taken as a superposition of the 
restricted GWPs \eqref{eq_res_gwp}.
These basis states are then propagated analytically in time.
We start with the derivation of the time evolution of the basis states.
Inserting the ansatz \eqref{eq_gwpKS} in the time-dependent Schr\"odinger 
equation \eqref{eq_regH} yields
\begin{eqnarray}
 &&\left(i\frac{\partial}{\partial\tau}-H\right) g({\bf y},\x) \\
 &=& \left(-\dot\gamma+i\,{\rm tr}\, A-\uu(\dot A+2A^2)\uu-\frac{1}{2}\uu^2
   \right) g({\bf y},\x) = 0 \; . \nonumber
\label{eq_schr}
\end{eqnarray}
The equations of motion for the $4\times 4$ width matrix $A$ given in
Eq.\ \eqref{eq_A_sym} and the complex phase factor $\gamma$ are now
directly obtained from Eq.\ \eqref{eq_schr} as
\begin{subequations}
\begin{align}
 \dot A & =  -2 A^2 -\frac{1}{2} {\boldsymbol 1} \; ,
\label{eq_A_dot} \\ 
 \dot \gamma & =  i\,{\rm tr}\, A \; .
\label{eq_g_dot}
\end{align}
\label{eq_Ag_dot}
\end{subequations}
It is important to note that although the parameter set of the GWP
\eqref{eq_gwpKS} is restricted to only five complex parameters 
(as compared to 15 parameters in \eqref{GWPhel1})
the differential equations \eqref{eq_Ag_dot} still describe the exact
dynamics of the wave packet without any approximation, i.e., with the
initial condition \eqref{eq_nb_0} the wave packet stays centered around 
the origin for all times, and the width matrix $A$ keeps the form 
\eqref{eq_A_sym} because the matrix $\dot A$ in \eqref{eq_A_dot} has 
the same structure \eqref{eq_A_sym} as $A$ itself.

Both equations \eqref{eq_Ag_dot} can be solved analytically.
Equation \eqref{eq_A_dot} is solved most easily when two 
auxiliary complex matrices $B$ and $C$ are introduced 
with $A= \frac{1}{2}BC^{-1}$, 
 and the initial conditions $B(0) = 2A(0)$  
and $C(0) = {\bf 1}$ \cite{Hel76a}.
Then \eqref{eq_A_dot} is replaced with the two equations $\dot C = B$ and 
$\dot B =- C$, or equivalently $\ddot B= -B$.
The matrices $B$ and $C$ have the same structure as the width matrix $A$ 
\eqref{eq_A_sym}, with the solution for the matrix $B$ 
\begin{eqnarray}
 b_\mu(\tau) & = & 2 a_\mu^0 \cos \tau -\sin \tau \; , \nonumber\\
 b_\nu(\tau) & = & 2 a_\nu^0 \cos \tau -\sin \tau \; ,\nonumber \\
 b_x  (\tau) & = & 2 a_x^0 \cos \tau \; ,\nonumber \\
 b_y  (\tau) & = & 2 a_y^0 \cos \tau \; ,
\end{eqnarray}
and for the matrix C
\begin{eqnarray}
 c_\mu (\tau)& =& \cos \tau +2 a_\mu^0 \sin \tau \; ,\nonumber\\
 c_\nu (\tau)& = & \cos \tau +2 a_\nu^0 \sin \tau \; ,\nonumber \\
 c_x   (\tau)& =& 2a_x^0 \sin \tau \; ,\nonumber \\
 c_y (\tau)  & = & 2a_x^0 \sin \tau \; .
\end{eqnarray}
Here and in the following the superscript $0$ indicates parameters of the
initial state at time $\tau=0$.
The matrix $A$ is obtained from the above definition $A =\frac{1}{2} B C^{-1}$ 
and the four elements in the form of Eq.\ \eqref{eq_aparam_cart} read
\begin{eqnarray}
 p_r(\tau) & = & \frac{1}{Z(\tau)}
 \left\{p_r^0 \cos 2\tau + \frac{1}{2} \left[(p_r^0)^2-(\p^0)^2 -1\right]
   \sin 2\tau\right\} , \nonumber\\ 
 p_x(\tau) & = & \frac{p_x^0}{Z(\tau)} \; , \quad %\nonumber\\ 
 p_y(\tau) = \frac{p_y^0}{Z(\tau)} \; , \quad %\nonumber\\ 
 p_z(\tau) = \frac{p_z^0}{Z(\tau)} \; ,
\label{eq_pi_tau}
\end{eqnarray}
where $Z(\tau)$ abbreviates the expression 
\begin{equation}
 Z(\tau) = \cos^2 \tau + \left[(p_r^0)^2-(\p^0)^2\right] \sin^2 \tau
 + p_r^0 \sin2\tau \; .
\end{equation}
With the matrix $A$ at hand it is possible to integrate Eq.\ 
\eqref{eq_g_dot} to obtain $\gamma$. The quantity $e^{-i\gamma}$, i.e.,  
the phase and normalization of the wave function, then reads, with the 
initial value $\gamma(0)=0$
\begin{equation}
 {\mathcal N}(\tau) \equiv e^{-i\gamma(\tau)}
 = 1 + \left[(p_r^0)^2-(\p^0)^2\right](1-\cos2\tau) + p_r^0\sin2\tau .
\label{eq_g_tau}
\end{equation}
The analytical time evolution of the wave function is obtained by inserting 
the time-dependent parameters in the wave function \eqref{eq_KS_gwp_x} 
which finally yields
\begin{widetext}
\begin{equation}
 g(\tau,{\bf y}^0,\x)  = \frac{1}{\mathcal N(\tau)} 
 \exp\left\{i\,\frac{2(\p^0 \cdot \x +p_r^0 r\cos2\tau)
  +[(p_r^0)^2-(\p^0)^2-1]r\sin2\tau}{2\mathcal N(\tau)} \right\} \; .
\label{eq_GWP_tau}
\end{equation}
\end{widetext}
This is an important intermediate result.
The time evolution of a restricted GWP \eqref{eq_KS_gwp_x} has been
 calculated analytically 
 and takes the compact form \eqref{eq_GWP_tau}.
The parameters in Eqs.\ \eqref{eq_pi_tau} and  \eqref{eq_g_tau} are periodic 
functions of the time $\tau$ with period $\pi$.
This results in a  $\pi$-periodicity of the  wave function 
\eqref{eq_GWP_tau}.
In the physical time, wave packets disperse in the hydrogen atom 
\cite{Bhaumik86,Str93}. 
 By contrast, the wave packets in the fictitious time show an oscillating 
behavior with no long-time dispersion in $\tau$.

Now that the time evolution of the basis states \eqref{eq_KS_gwp_x} is known we 
can expand an arbitrary initial state in this basis.
The time evolution of that wave function is then analytically given by the 
superposition of the time-dependent restricted GWPs \eqref{eq_GWP_tau}.
In general, the expansion of an arbitrary state $\psi({\bf x})$ in an 
over-complete set of Gaussian wave functions $g({\bf y},{\bf x})$ is
a nontrivial task.
A procedure for finding the ``optimal'' expansion for a given number $N$ of 
basis states is to minimize the deviation
$ \Delta = ||\psi(\x)-\sum_{k=1}^Ng({\bf y}^k,\x)||^2 $,
e.g., by searching for stationary points 
$\frac{\partial\Delta}{\partial{\bf y}^k}=0$ with respect to the 
parameters ${\bf y}^k,\;k=1,\dots,N$ \cite{Saw85}. 
This procedure presents a highly nonlinear minimization problem.
Many stationary points, i.e., local minima may exist, and the difficulty 
is to find the true global minimum.

Here, we concentrate on the propagation of 3D Gaussian wave packets which 
are localized around a point $\x_0$ with width $\sigma$ in coordinate space 
and around $\p_0$ in momentum space, and present a direct approach for 
the expansion of the 3D GWPs given as
\begin{equation}
 \psi(\x) = (2\pi \sigma^2)^{-3/4}
 \exp\left\{-\frac{(\x-\x_0)^2}{4\sigma^2}+i\,\p_0\cdot(\x-\x_0)\right\} 
\label{eq_psi_ini_pos}
\end{equation}
in the restricted GWPs \eqref{eq_res_gwp}.
The Fourier representation of $\psi(\x)$ reads 
\begin{eqnarray}
     \psi(\x)
 &=& \left(\frac{\sigma^2}{2\pi^3}\right)^{3/4}
      \int d^3p\,e^{-\sigma^2(\p-\p_0)^2+i\,\p\cdot (\x-\x_0)} \; .
\label{invFTpsi}
\end{eqnarray}
Using the approximation $r\approx\x\cdot\x_0/|\x_0|$,
which is valid in the vicinity of $\x_0$ where $\psi({\bf x})$ is localized 
Eq.\ \eqref{invFTpsi} can be written as
\begin{eqnarray}
     \psi(\x)
 &\approx& \left(\frac{\sigma^2}{2\pi^3}\right)^{3/4}
     \int d^3p\,e^{-\sigma^2(\p-\p_0)^2-i\,\p\cdot\x_0} \nonumber \\
 &\times& e^{i[p_r r+(\p-p_r\frac{\x_0}{|\x_0|})\cdot\x]} \\
 &=& \left(\frac{\sigma^2}{2\pi^3}\right)^{3/4}
     \int d^3p\,e^{-\sigma^2(\p-\p_0)^2-i\,\p\cdot\x_0}\,
     g({\bf y},\x) \; , \nonumber
\label{invFTpsi2}
\end{eqnarray}
where the $g({\bf y},\x)$ are the restricted GWPs \eqref{eq_KS_gwp_x} for 
the set of parameters ${\bf y}$ given by
$(p_r=\textrm{const.},\p-p_r\x_0/|\x_0|,\gamma=0)$.
As can be easily shown the restricted GWPs \eqref{eq_KS_gwp_x} for 
constant $p_r$ and $\gamma=0$ are a complete continuous 
basis as functions of the momentum ${\bf p}$, i.e.
\begin{equation}
 \frac{1}{(2\pi)^3} \int d^3p\, |g({\bf y})\rangle\langle g({\bf y})|
 = {\boldsymbol 1} \; .
\label{compl_rel}
\end{equation}
Note that the completeness relation \eqref{id_GWP} is valid for frozen 
Gaussians with arbitrary localization in coordinate and momentum space 
while in Eq.\ \eqref{compl_rel} the restricted GWPs in KS coordinates are 
all located around the origin but the width matrix $A$ is varied.
Note also that Eq.\ \eqref{invFTpsi2} is exact for $p_r=0$, i.e., for 
the restricted GWPs given as plane waves, and becomes an approximation 
for $p_r\ne 0$.
This means that $p_r$ should be chosen as zero or close to zero in practical
applications.

In numerical computations it is convenient to approximate the 3D Gaussian 
wave packet \eqref{eq_psi_ini_pos} by a finite number of restricted GWPs
rather than using the integral representation \eqref{invFTpsi2}.
This is most efficiently achieved by evaluating the integral in 
\eqref{invFTpsi2} with a Monte Carlo method using importance sampling of 
the momenta with a normalized Gaussian weight function
\begin{equation}
 w(\p)= \left(\frac{\sigma^2}{\pi}\right)^{3/2} e^{-\sigma^2(\p-\p_0)^2} \; .
\label{eq_weight}
\end{equation}
The initial wave packet then reads
\begin{equation}
 \psi(\x) = (2\pi\sigma^2)^{-3/4} \frac{1}{N}
 \sum_{k=1}^N g\left({\bf y}^k,\x\right) \, e^{-i\,\p^k\cdot\x_0} \; ,
\label{psi_MC}
\end{equation}
with ${\bf y}^k=(p_r,\p^k-p_r\x_0/|\x_0|,0)$, and the $\p^k$, $k=1,\dots,N$ 
distributed randomly according to the normalized Gaussian weight function 
\eqref{eq_weight}.
The wave function $\psi(\x)$ in Eq.\ \eqref{psi_MC} is an approximation
to the 3D Gaussian wave packet \eqref{eq_psi_ini_pos}, and the accuracy
depends on how many restricted GWPs are included.
However, it is important to note that a wave packet which is strongly 
localized around $\x_0$ in coordinate space and around $\p_0$ in momentum 
space can be described even with a rather low number $N$ (of the order of 
50-100) restricted GWPs.

The time propagation of an initial state \eqref{psi_MC} in the fictitious
time $\tau$ is now obtained exactly and fully analytically in a simple way.
In the unperturbed Coulomb problem a superposition of time-dependent 
restricted GWPs is uncoupled, i.e., all basis functions propagate 
independently.
Thus, the time propagation of the initial state \eqref{psi_MC} simply reads
\begin{equation}
 \psi(\tau,\x) = (2\pi\sigma^2)^{-3/4} \frac{1}{N}
 \sum_{k=1}^N g\left(\tau,{\bf y}^k,\x\right) \, e^{-i\,\p^k\cdot\x_0} \; ,
\label{psi_tau}
\end{equation}
with the same parameters ${\bf y}^k$ as above and the time propagation
of the restricted GWPs given by Eq.\ \eqref{eq_GWP_tau}.

\begin{figure*}
%\begin{widetext}
\includegraphics[width=0.95\textwidth]{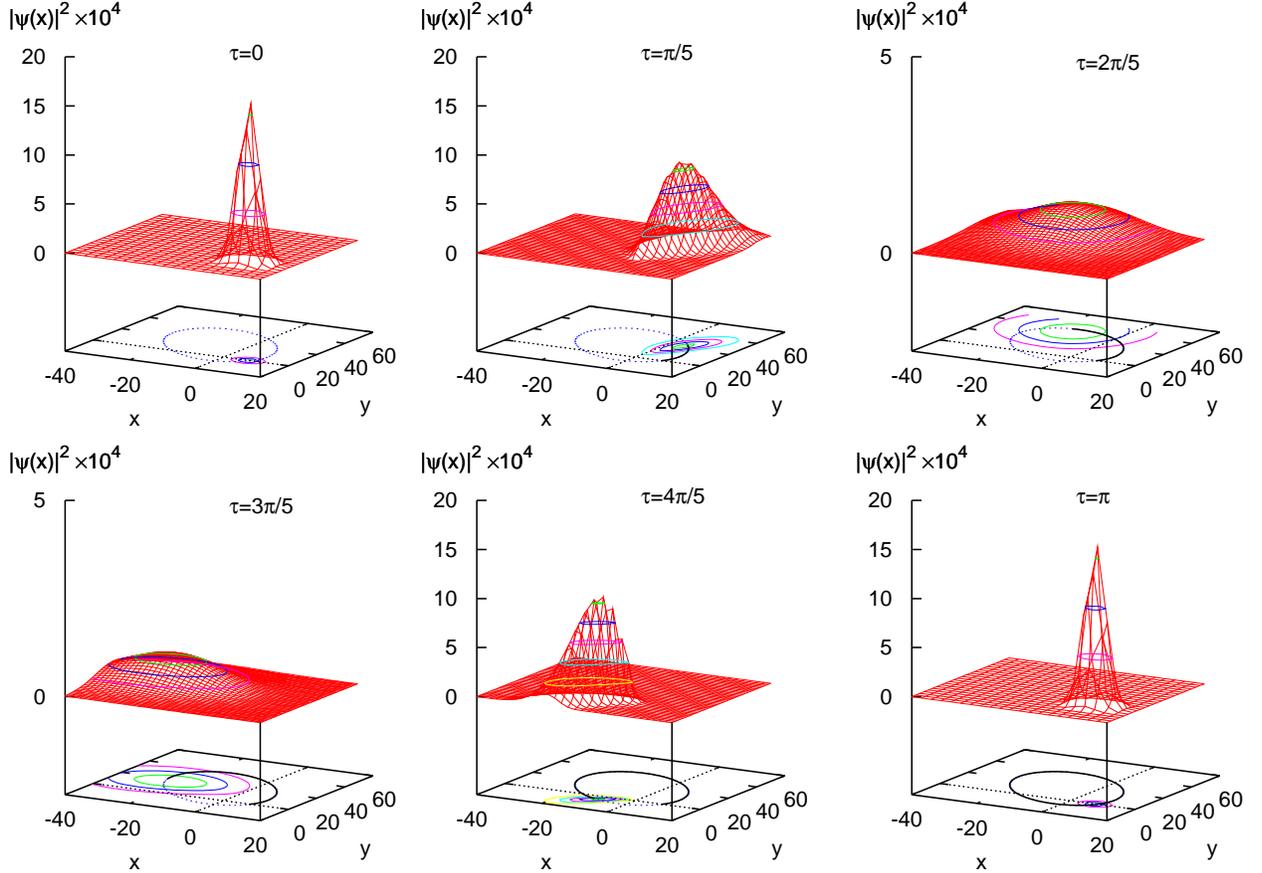}
\caption{(Color online) Fictious time propagation of the initially Gaussian
  wave packet
  \eqref{eq_psi_ini_pos} located at $\x_0=(8,0,0)$ with the momentum
  $\p_0=(1,2,0)$ plotted in the plane $z=0$.  The classical Kepler ellipse
  with the same initial conditions is indicated by the dotted curve
  on the bottom
  of each panel.  For time resolved comparison, that part of the ellipse 
  that has been traversed by the particle so far in each plot is shown by a
  solid black line.  Although the Gaussian wave packet does not
  stay Gaussian during the period  it follows in general the classical path
  and is recovered after one period $\tau = \pi$, indicating the periodicity
  of the wave packet.}
\label{fig_psi_mc_3d_ellipse}
%\end{widetext}
\end{figure*}

Eq.\ \eqref{psi_tau} is the final result of this section, and is illustrated
in Fig.\ \ref{fig_psi_mc_3d_ellipse} for an initially Gaussian wave packet 
\eqref{eq_psi_ini_pos} whose center is moving in the $z=0$ plane.
The wave packet is expanded and propagated analytically in a basis 
of $N=10000$ restricted GWPs. 
 This is possible since the analytical approach allows for a large
number of basis states.
We note, however, that results of similar quality can be obtained using $N=50\sim 100$
 GWPs only. 
%
%\onecolumngrid
%
%\twocolumngrid
The probability density in the $z=0$ plane is plotted at equidistant times 
with step size $\Delta\tau = \pi/5$.
As mentioned above Eq.\ \eqref{invFTpsi2} is exact for $p_r=0$, however, in 
our numerical calculations we choose $p_r=i\epsilon$ with a small $\epsilon>0$.
The damping enables normalization of the restricted GWPs, and improves the
convergence of the Monte Carlo integral.
The initial GWP presented in Fig.\ \ref{fig_psi_mc_3d_ellipse} for 
$\tau = 0$ is centered at $\x_0 = (8,0,0)$ with the mean momentum
$\p_0 = (1,2,0)$, and $p_r$ is set to $p_r=i\epsilon$ with $\epsilon = 0.01$.
Classically the electron with these initial conditions is running on the 
Kepler ellipse plotted by dots on the bottom of each panel in 
Fig.\ \ref{fig_psi_mc_3d_ellipse}.
For every time step, that part of the ellipse that has been passed by the 
electron so far is shown by a black solid line for time resolved comparison.
The position of the maximum of the probability density agrees well with the 
classical position of the electron on the ellipse for all times.
The $\pi$-periodicity of the motion is reflected by the coincidence of the 
wave packet after one period at $\tau = \pi$ with the initial GWP at 
$\tau = 0$. 

\subsection{Propagation of 2D cylindrically symmetric wave packets}
\label{sec_g_lz_m}
In this section basis functions based on the restricted GWP \eqref{eq_KS_gwp} 
with a well defined  angular momentum component $l_z=m$ are derived. 
This case is especially important when a cylindrically symmetric external
field, e.g.\ a magnetic field, is applied to the hydrogen atom, 
as discussed in the following paper \cite{Fab08b}.
First wave packets with definite $l_z$ are constructed and their exact, 
analytical dynamics in the hydrogen atom is discussed.
Then we introduce a procedure to expand quantum states of defined $l_z$ in 
terms of the basis states.  

A 2D cylindrically symmetric restricted GWP \eqref{eq_res_gwp} is obtained 
by setting $a_x=a_y=0$ in \eqref{eq_KS_gwp} and introducing parabolic
coordinates $\xi=\mu^2=r+z$, $\eta=\nu^2=r-z$, i.e.\ 
\begin{equation}
 g_0({\bf y},\x) = e^{i(a_\mu \mu^2+a_\nu \nu^2 +\gamma)}
                 = e^{i(p_\xi \xi+ p_\eta \eta + \gamma)} \; ,
\label{eq_KS_gwp_x_lz_0}
\end{equation}
with the parabolic momenta $p_\xi=a_\mu$ and $p_\eta=a_\nu$.
These states are axisymmetric, and have the quantum number $m=0$, but can be 
generalized to arbitrary $m$ by setting
\begin{eqnarray}
 g_m({\bf y},\x) &=& (\mu \nu )^{|m|}e^{i(a_\mu \mu^2+a_\nu \nu^2 +\gamma)} \,
 e^{i m \varphi} \nonumber \\
 &=& (\xi\eta)^{|m|/2} e^{i(p_\xi \xi+ p_\eta \eta + \gamma)}\, e^{i m \varphi} \; .
\label{eq_KS_gwp_x_lz_m}
\end{eqnarray}
The time-dependent parameters are ${\bf y} = (a_\mu,a_\nu,\gamma)$
or equivalently ${\bf y} = (p_\xi,p_\eta,\gamma)$.
The quantum number $m$ is constant. 
As will be shown, the wave packet \eqref{eq_KS_gwp_x_lz_m} still presents 
an exact solution of the regularized Schr\"odinger equation.
The Laplacian in semiparabolic coordinates \eqref{eq_Laplace_op_semi} 
and the time derivative acting on the wave packet \eqref{eq_KS_gwp_x_lz_m} yield
\begin{widetext}
\begin{eqnarray}
 \Delta g_m({\bf y},\x) & = & 
 \left[4i\left(a_\mu+a_\nu \right) \left( 1 + |m| \right) 
   - 4a_\mu^2 \mu^2 - 4 a_\nu^2 \nu^2\right] g_m({\bf y},\x) \; , \nonumber \\
 i \frac{\partial}{\partial \tau}g_m({\bf y},\x)
 & = &  \left( - \dot a_\mu \mu^2 - \dot a_\nu \nu^2 -\dot \gamma \right)
 g_m({\bf y},\x) \; ,
\label{eq_ipsid_mTpsi_m}
\end{eqnarray}
and thus the time-dependent Schr\"odinger equation reads
$[(\dot a_\mu + 2a_\mu\mu^2+\frac{1}{2})\mu^2+(\dot a_\nu+2a_\nu\nu^2
+\frac{1}{2})\nu^2+\dot\gamma-2i(a_\mu+a_\nu)(1+|m|)]\, g_m({\bf y},\x)=0$.
This equation is solved exactly if the Gaussian parameters obey the equations 
of motion
\begin{subequations}\label{eq_dot_lz_m_comp}
\begin{align}
 \dot a_\mu & =  -2 a_\mu^2-\frac{1}{2} \; , \\
 \dot a_\nu & =  -2 a_\nu^2-\frac{1}{2} \; , \\
 \dot \gamma & = 2i \left(a_\mu + a_\nu \right) \left( 1+ |m| \right) \; ,
\end{align}
\end{subequations} 
or using the matrix notation \eqref{eq_A_sym} for $A$ (with $a_x=a_y=0$)
\begin{subequations}
\label{eq_dot_lz_m}
\begin{align}
\dot A  & =  -2 A^2 - \frac{1}{2} {\boldsymbol 1} \; , \label{eq_A_dot_lz_m}\\
\dot \gamma & =  i\, {\rm tr}\, A \left( 1+ |m| \right) \; . 
\label{eq_g_dot_lz_m}
\end{align}
\end{subequations}
The equations of motion for the two nonzero complex width parameters $a_\mu$ 
and $a_\nu$ remain completely unchanged as compared to the restricted GWP in 
Sec.\ \ref{sec_expansion_res}.
The only change is the additional factor of $\left( 1+ |m| \right)$ in 
Eq.\ \eqref{eq_g_dot_lz_m} for the phase parameter $\gamma$.
The solution of Eq.\ \eqref{eq_A_dot_lz_m} is
\begin{eqnarray}
 a_\mu(\tau) & = & \frac{1}{Z(\tau)} \left\{2(a_\mu^0-a_\nu^0)+2(a_\mu^0+a_\nu^0)
 \cos 2\tau - (1- 4 a_\mu^0 a_\nu^0)\sin 2\tau \right\} \; , \nonumber \\
 a_\nu(\tau) & = & \frac{1}{Z(\tau)} \left\{2(a_\nu^0-a_\mu^0)+2(a_\mu^0+a_\nu^0)
 \cos 2\tau - (1- 4 a_\mu^0 a_\nu^0)\sin 2\tau \right\} \; , %\nonumber 
\end{eqnarray}
with
\begin{equation}
 Z(\tau) = 2 \left[ 1+ 4 a_\mu^0 a_\nu^0 +
  \left( 1 - 4 a_\mu^0 a_\nu^0 \right) \cos 2\tau  
  + 2 \left( a_\mu^0 + a_\nu^0 \right) \sin 2\tau \right] \; .
\end{equation}
%\end{widetext}
%
The solution of Eq.\ \eqref{eq_g_dot_lz_m} is the solution of 
Eq.\ \eqref{eq_g_dot} multiplied by the factor $(1+|m|)$, 
and the phase and normalization factor of the wave function 
 reads,  with $\gamma(0) = 0$
\begin{equation}
e^{-i\g} = \left(\frac{1}{4}Z(\tau)\right)^{|m|+1}\;.
\end{equation}
The time evolution of the wave packet \eqref{eq_KS_gwp_x_lz_m} then is given by
\begin{eqnarray}
 g_m(\tau,\y^0,\mu,\nu) 
 &=& \frac{1}{\left(\frac{1}{4}Z(\tau)\right)^{|m|+1}}(\mu \nu)^{|m|} \,
 e^{i[a_\mu(\tau)\mu^2+a_\nu(\tau)\nu^2]} \, e^{i m \varphi} \; .
\label{eq_gm_tau}
\end{eqnarray}
The time-dependent basis states \eqref{eq_gm_tau} are the analogue of
the restricted GWPs \eqref{eq_GWP_tau} for the propagation of wave packets 
with constant magnetic quantum number $m$.
They obey the constraint \eqref{KScon}, but they are not sufficiently 
general to describe the dynamics of 2D wave packets localized around a
given point $(\xi_0,\eta_0,p_{\xi_0},p_{\eta_0})$ in the parabolic coordinate
phase space.
Such localized states can now be constructed in a similar way as described
in Sec.\ \ref{sec_expansion_res}.
We use the formal plane wave expansion of a Gaussian wave packet
in parabolic coordinates, viz.\  
%
%\begin{widetext}
\begin{eqnarray}
     \psi_0(\xi,\eta)
 &=&{\mathcal A} \exp\left\{-\frac{(\xi-\xi_0)^2}{4\sigma^2}
     - \frac{(\eta-\eta_0)^2}{4\sigma^2}
     + ip_{\xi_0}(\xi-\xi_0) + ip_{\eta_0}(\eta-\eta_0)\right\} \nonumber \\
 &=&\frac{{\mathcal A}\sigma^2}{\pi} \int dp_\xi \int dp_\eta\,
     e^{-\sigma^2[(p_\xi-p_{\xi_0})^2+(p_\eta-p_{\eta_0})^2]
      - i(p_\xi\xi_0+p_\eta\eta_0)} \,
     e^{i(p_\xi\xi+p_\eta\eta)} \nonumber \\
 &=&\frac{{\mathcal A}\sigma^2}{\pi} \int dp_\xi \int dp_\eta\,
     e^{-\sigma^2[(p_\xi+i\epsilon-p_{\xi_0})^2+(p_\eta+i\epsilon-p_{\eta_0})^2]
      + \epsilon(\xi_0+\eta_0) - i(p_\xi\xi_0+p_\eta\eta_0)}\,
     e^{i[(p_\xi+i\epsilon)\xi+(p_\eta+i\epsilon)\eta)]} \nonumber \\
 &=&\frac{{\mathcal A}\sigma^2}{\pi} \int dp_\xi \int dp_\eta\,
     e^{-\sigma^2[(p_\xi+i\epsilon-p_{\xi_0})^2+(p_\eta+i\epsilon-p_{\eta_0})^2]
     + \epsilon(\xi_0+\eta_0) - i(p_\xi\xi_0+p_\eta\eta_0)}\, g_0({\bf y},\x)\; ,
\label{eq_GWP_parabol}
\end{eqnarray}
%\end{widetext}
%
where ${\mathcal A}$ is a normalization factor, $\epsilon$ has been introduced 
(without any approximation) as an
additional free parameter, and $g_0({\bf y},\x)$ is the cylindrically symmetric
restricted GWP \eqref{eq_KS_gwp_x_lz_0} for the set of parameters
${\bf y} = (p_\xi+i\epsilon,p_\eta+i\epsilon,\gamma=0)$.
A value of $\epsilon>0$ guarantees that $g_0({\bf y},\x)$ can be normalized.
From \eqref{eq_GWP_parabol} an initial state with given magnetic quantum
number $m$ is obtained as
\begin{equation}
 \psi_m(\x) = (\xi\eta)^{|m|/2} \psi_0(\xi,\eta)\, e^{i m \varphi}\;.
\end{equation}
In numerical computations the integrals in \eqref{eq_GWP_parabol} are
approximated employing a Monte Carlo technique in the same way as explained in 
Sec.\ \ref{sec_expansion_res}.
We obtain
\begin{eqnarray}
 \psi_m(\x) &\approx&\frac{{\mathcal A}\sigma^2}{\pi}
 (\xi\eta)^{|m|/2} \frac{1}{N}\sum_{k=1}^N g_0({\bf y}^k,\x) \,
 e^{-i(p^k_\xi\xi_0+p^k_\eta\eta_0)+\epsilon(\xi_0+\eta_0)
   -2i\sigma^2\epsilon[(p^k_\xi-p_{\xi_0})+(p^k_\eta-p_{\eta_0})]
   +2\sigma^2\epsilon^2}\, e^{i m \varphi} \nonumber \\
 &=&\frac{{\mathcal A}\sigma^2}{\pi} \frac{1}{N}\sum_{k=1}^N g_m({\bf y}^k,\x) \,
 e^{-i(p^k_\xi\xi_0+p^k_\eta\eta_0)+\epsilon(\xi_0+\eta_0)
   -2i\sigma^2\epsilon[(p^k_\xi-p_{\xi_0})+(p^k_\eta-p_{\eta_0})]
   +2\sigma^2\epsilon^2} \; ,
\end{eqnarray}
with sampling points $p_\xi^k$, $p_\eta^k$ randomly distributed around 
$p_{\xi_0}$, $p_{\eta_0}$ according to the weight function 
$w(p) = (\sigma/\sqrt{\pi})e^{-\sigma^2(p-p_0)^2}$.
Finally, the replacement of the initial basis states $g_m({\bf y}^k,\x)$
with the corresponding time-dependent solutions \eqref{eq_gm_tau} yields
\begin{eqnarray}
 \psi_m(\tau,\x) =\frac{{\mathcal A}\sigma^2}{\pi}
 \frac{1}{N}\sum_{k=1}^N g_m(\tau,{\bf y}^k,\x) \,
 e^{-i(p^k_\xi\xi_0+p^k_\eta\eta_0)+\epsilon(\xi_0+\eta_0)
   -2i\sigma^2\epsilon[(p^k_\xi-p_{\xi_0})+(p^k_\eta-p_{\eta_0})]
   +2\sigma^2\epsilon^2} \; ,
\end{eqnarray}
\end{widetext}
with the parameter sets 
${\bf y}^k = (p^k_\xi+i\epsilon,p^k_\eta+i\epsilon,\gamma=0)$.

\begin{figure*}
\includegraphics[width = 0.95\textwidth]{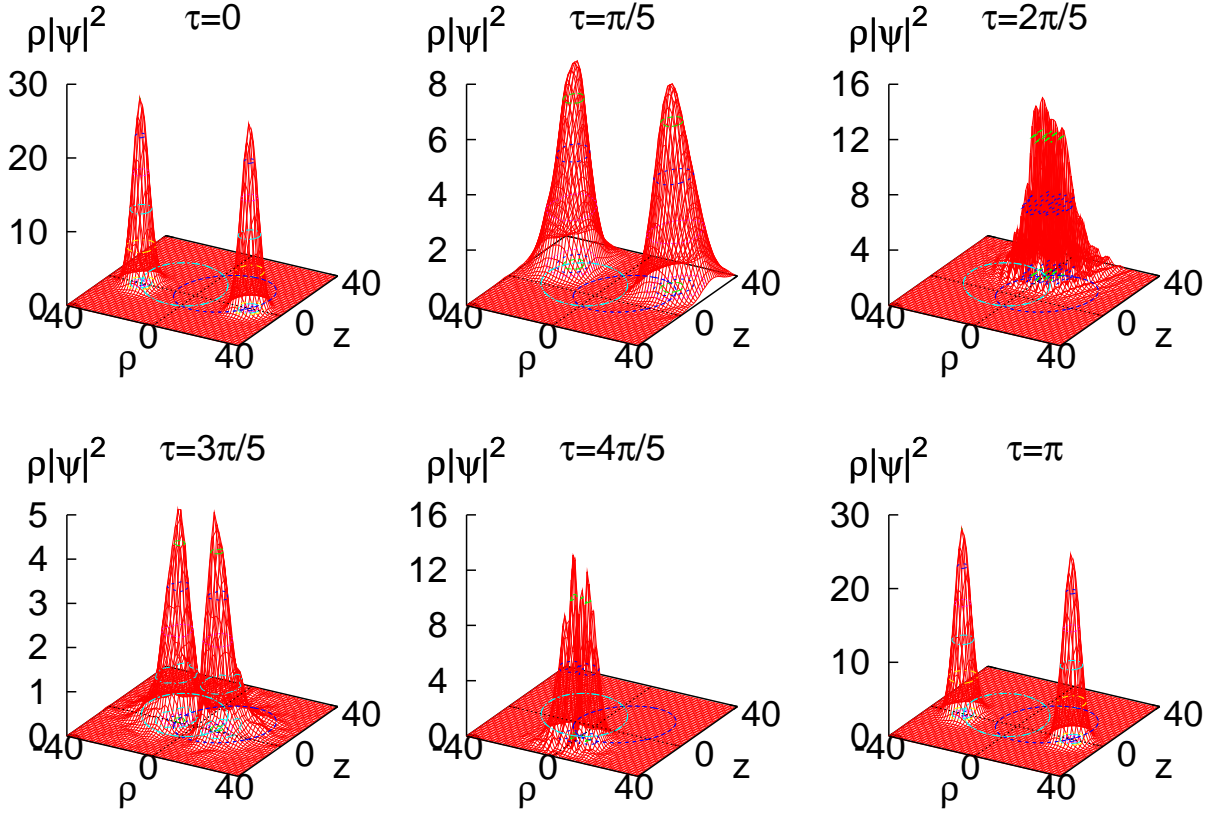}
\caption{(Color online) 
 Fictitious time propagation of a GWP \eqref{eq_GWP_parabol} with magnetic quantum
  number $m=0$.  The part on the negative $\rho$ axis is obtained by
  reflecting the positive part at the $z$ axis.  
  The wave packet runs along the classical Kepler ellipse with
  the corresponding initial values. For details see text.
\label{fig_psi_lz_0}}
\end{figure*}

In Fig.\ \ref{fig_psi_lz_0} the expansion of a GWP \eqref{eq_GWP_parabol} 
is shown with the center at $\xi_0=\eta_0 = 25.0$ and the momenta 
$p_{\xi_0}= 0.535$ and $p_{\eta_0}=-0.117$ or in terms of cylindrical 
coordinates $\rho_0=25.0, z_0 = 0.0$ and $p_{\rho_0}=0.419,p_{z_0}=0.652$ 
and the width $\sigma = 4.472$. 
Note that a Gaussian shape of a wave packet in parabolic coordinates is 
nearly Gaussian also in cylindrical coordinates (see e.g.\ the wave packet 
at $\tau=0$ in Fig.\ \ref{fig_psi_lz_0}).
The wave function shown has zero angular momentum component $l_z=0$.
For reasons of presentation the originally positive radial 
coordinate $\rho$ is extended to 
negative values, and the symmetry $\psi(-\rho) = \psi(\rho)$ is used.
 The probability density $\rho |\psi(\rho,z)|^2$ is plotted in 
the $(\rho,z)$ plane. 
The Kepler ellipses plotted on the bottom in each panel of 
Fig.\ \ref{fig_psi_lz_0} show the corresponding classical motion of the 
particle with the initial conditions given above.
The second ellipse again is obtained by reflection symmetry as the
intersection of the torus, which is obtained from rotating the ellipse 
around the $z$ axis.
At times $\tau \approx 2 \pi /5$ and $\tau \approx 4 \pi /5$ the high 
probability density close to the $z$ axis leads to interference patterns.
After one period $\tau = \pi$ the initial wave function $\tau=0$ is recovered.
 A number of $N=5000$ modified basis states \eqref{eq_KS_gwp_x_lz_m} with 
$\epsilon =0.05$ are employed. 
Results for the case $m \ne 0$ are not shown since they differ only qualitatively 
by avoiding the crossing of the $z$ axis due to the rotational barrier.

\subsection{Propagation of 1D spherically symmetric wave packets}
\label{sec_radial_gwp}
The procedure of the two previous subsections is applied to quantum states 
with conserved angular momentum.
First an extension of the basis states \eqref{eq_KS_gwp_x} to basis states 
with well defined angular momentum quantum numbers $lm$ is presented, and 
they are shown to be exact solutions of the time-dependent Schr\"odinger 
equation of the regularized hydrogen atom.  
Then the procedure of expanding states with definite $lm$ in the constructed 
basis states together with an example are presented. 
For radial symmetry the complex width matrix $A$ \eqref{eq_A_sym} of the 
restricted GWP must be a multiple of the identity matrix 
$A = a{\boldsymbol 1}$, i.e.\ $a_x=a_y=0$, $a_\mu=a_\nu\equiv a=p_r/2$.
The restricted GWP reduces to
\begin{equation}
 g_{00}(r) = e^{i(2ar+\gamma)} =  e^{i(p_r r+\gamma)} \; .
\label{g00}
\end{equation}
This is a suitable basis state with vanishing angular momentum.
The correct extension to arbitrary angular momenta is given by 
\begin{equation}
 g_{lm}(r,\theta,\varphi) = r^l e^{i(2ar+\gamma)} Y_{lm}(\theta,\varphi) \; ,
\label{eq_KS_gwp_x_l_m_kugel}
\end{equation}
where $Y_{lm}(\theta,\varphi)$ denotes the spherical harmonics.
Insertion of the ansatz \eqref{eq_KS_gwp_x_l_m_kugel} into the time-dependent 
Schr\"odinger equation \eqref{H_KSreg} in spherical coordinates
\begin{equation}
 i \frac{\partial}{\partial\tau} g_{lm}(r,\theta,\varphi) =
 \left(-\frac{\partial^2}{\partial r^2}r + \frac{l^2}{r} + r \right)
 g_{lm}(r,\theta,\varphi)
\end{equation} 
yields
\begin{equation}
 \left[-4i(l+1)a+\dot \gamma +r(1+(2a)^2+2\dot a) \right]
 g_{lm}(r,\theta,\varphi) = 0 \; .
\end{equation}
The basis sates \eqref{eq_KS_gwp_x_l_m_kugel} present an exact solution of 
the Schr\"odinger equation provided the time-dependent parameters obey the 
equations of motion 
\begin{subequations}
\begin{align}
 \dot a & = -2 a^2- \frac{1}{2}, \\
 \dot \gamma & = 4ia (l+1),
\end{align}
\end{subequations}
with the analytic solution
\begin{equation}
 a(\tau) = \frac{4a^0\cos 2\tau-(1-4(a^0)^2)\sin 2\tau}
  {2\left[1+4(a^0)^2+(1-4(a^0)^2) \cos 2\tau + 4 a^0 \sin 2\tau\right]}
\label{eq_a_tau}
\end{equation}
and
\begin{eqnarray}
 && e^{-i\gamma(\tau)} \\
 &=& \left[1+4(a^0)^2+(1-4(a^0)^2)\cos 2\tau+4a^0\sin 2\tau \right]/2 \; .
  \nonumber
\label{eq_gamma_tau}
\end{eqnarray}
Wave packets with well defined angular momentum quantum numbers $l$ and $m$
can now be expanded in the basis \eqref{eq_KS_gwp_x_l_m_kugel}.
For the radial coordinate $r$ the same procedure as introduced in Sec.\
\ref{sec_g_lz_m} for the parabolic coordinates $\xi$ and $\eta$ can be applied.
The plane wave expansion of a Gaussian wave packet localized around 
$(r_0,p_{r_0})$ reads
\begin{eqnarray}
     \psi_{00}(r)
 &=&{\mathcal A} \exp\left\{-\frac{(r-r_0)^2}{4\sigma^2}
     + ip_{r_0}(r-r_0) \right\} \\
 &=&\frac{{\mathcal A} \sigma}{\sqrt{\pi}} \int dp_r\,
     e^{-\sigma^2(p_r+i\epsilon-p_{r_0})^2
     + \epsilon r_0 - i p_r r_0}\, e^{i(p_r+i\epsilon)r} \nonumber \\
 &=&\frac{{\mathcal A} \sigma}{\sqrt{\pi}}  \int dp_r\,
     e^{-\sigma^2(p_r+i\epsilon-p_{r_0})^2
     + \epsilon r_0 - i p_r r_0}\, g_{00}({\bf y},r)\; , \nonumber
\label{eq_psi_radial}
\end{eqnarray}
where $g_{00}({\bf y},r)$ is the spherically symmetric restricted GWP 
\eqref{g00} for the set of parameters
${\bf y} = (p_r+i\epsilon,\gamma=0)$.
An initial state with given angular momentum quantum numbers $l$ and $m$
is obtained as
\begin{equation}
 \psi_{lm}(\x) = r^l \psi_{00}(r)\, Y_{lm}(\theta,\varphi) \; .
\label{eq_psi_lm}
\end{equation}
Using the Monte Carlo evaluation of the integral in \eqref{eq_psi_radial}
and the replacement of the initial basis states \eqref{g00} with the 
corresponding time-dependent solutions 
$g_{00}(\tau,r)=\exp\{i[2a(\tau)r+\gamma(\tau)]\}$ 
(with the time-dependent parameters given in Eqs.\ \eqref{eq_a_tau}
and \eqref{eq_gamma_tau}) we finally obtain
\begin{eqnarray}
 \psi_{lm}(\tau,\x) &=& \frac{{\mathcal A} \sigma}{\sqrt{\pi}} %\dots
 \frac{1}{N}\sum_{k=1}^N g_{lm}(\tau,{\bf y}^k,\x) \nonumber \\
 &\times& e^{-i p^k_r r_0+\epsilon r_0 - 2i\sigma^2\epsilon(p^k_r-p_{r_0})
   +\sigma^2\epsilon^2} \; ,
\end{eqnarray}
with sampling points $p_r^k$ randomly distributed around 
$p_{r_0}$ according to the weight function 
$w(p) = (\sigma/\sqrt{\pi})e^{-\sigma^2(p-p_0)^2}$, and the parameter sets 
${\bf y}^k = (p^k_r+i\epsilon,\gamma=0)$.

\begin{figure}
\centerline{\includegraphics[width=0.95\columnwidth]{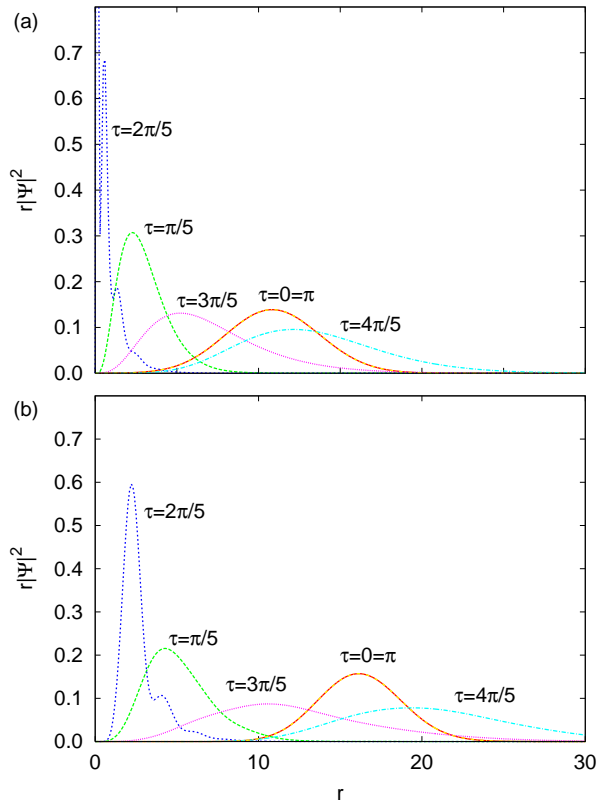}}
\caption{(Color online) 
  Fictitious time propagation of the wave function
  \eqref{eq_psi_lm} with $\psi(r)$
  given by the Gaussian \eqref{eq_psi_radial} with $r_0 = 10.0$,
  $ p_{r_0}=-0.5$ and $\sigma = 3.0$ for angular momentum quantum numbers
  (a) $l=0$, $m=0$, and (b) $l=5$, $m=0$.  An interference pattern is
  observed at the turning points at approximately $\tau = 2 \pi/5$ in both
  panels.  The initial wave function is expanded in $N=10000$ basis 
  states \eqref{eq_KS_gwp_x_l_m_kugel} with $\epsilon = 0.2$.}
\label{fig_psi_radial}
\end{figure}

The results of the propagation of the wave function 
$\psi_{lm}(r,\theta,\varphi)=r^l\psi_{00}(r)Y_{lm}(\theta,\varphi)$
with $\psi_{00}(r)$ given in Eq.\ \eqref{eq_psi_radial} and the initial 
values $r_0 = 10$, $p_{r_0}=-0.5$ and the width $\sigma = 3$ is presented
in Fig.\ \ref{fig_psi_radial} for different times $0 \le \tau \le \pi$.
The imaginary parts of $p_r^k=2a^k$ are set to $\epsilon = 0.2$ and the number 
of basis states \eqref{eq_KS_gwp_x_l_m_kugel} is $N=10000$. 
In Fig.\ \ref{fig_psi_radial}(a) the angular momentum is set to $m=l=0$ and in Fig.\  
\ref{fig_psi_radial}(b) the 
components of the angular momentum are $l=5$ and $m=0$.
Due to the negative initial value of the radial momentum $p_{r_0}$ the wave 
is initially running towards the nucleus located at the origin.
The wave function with zero angular momentum in Fig.\ \ref{fig_psi_radial}(a) 
comes close to the origin $r=0$.
Similar to the radial symmetric case in Sec.\ \ref{sec_g_lz_m} there appears 
to occur some interference pattern due to the overlapping parts of the 
incoming wave function at the inner turning point (see $\tau = 2 \pi /5$).
In the nonvanishing angular momentum case (Fig.\ \ref{fig_psi_radial}(b)) 
the barrier of rotational energy prevents the wave function from reaching 
the nucleus.
Instead there is a turning point whose distance from the nucleus increases 
with growing angular momentum. 
Again at $\tau = 2\pi/5$ an interference pattern is observed close to  
the inner turning point. 
In both panels the maximum of the probability density overshoots the position 
of the initial maximum $r_0=10$ at $\tau = 4 \pi /5$ due to the initial 
kinetic energy and returns to the initial wave packet after the period 
$\tau  = \pi$, indicating the periodicity of the wave function.

\section{Conclusion}
\label{sec:conclusion}
In this paper we have derived the wave packet dynamics for the field-free
hydrogen atom in a fictitious time variable.
The Coulomb problem has been transformed to the four-dimensional harmonic 
oscillator in Kustaanheimo-Stiefel coordinates with a constraint.
The ``restricted Gaussian wave packets'' obeying that condition have been 
constructed and their exact time dependence is calculated analytically.
The wave packets with and without symmetries exhibit a nondispersive
periodic behavior in the fictitious time.

It should be noted that the wave packet propagation in the fictitious time
substantially differs from the physical time dynamics, and thus cannot provide
the analytical propagation of dispersive wave packets in the physical time
\cite{Barnes93,Barnes94}.
Nevertheless, the fictitious time dynamics can be used to solve the
Schr\"odinger equation for Coulomb systems with strong time-independent 
perturbations, e.g., the hydrogen atom in static external electric and 
magnetic fields.
The restricted Gaussian wave packets are the basis for the application of 
the time-dependent variational principle to the hydrogen atom in external 
fields and the computation of quantum spectra by frequency analysis of 
the time autocorrelation function in the following paper \cite{Fab08b}.
As a consequence of using the fictitious time variable the method is exact 
for the field-free hydrogen atom and approximations in the variational 
approach are only induced by the external fields.

%\bibliography{paper}

\end{document}